\documentclass{article}

\usepackage{arxiv}

\usepackage[utf8]{inputenc} 
\usepackage[T1]{fontenc}    
\usepackage{hyperref}       
\usepackage{url}            
\usepackage{booktabs}       
\usepackage{amsfonts}       
\usepackage{nicefrac}       
\usepackage{microtype}      
\usepackage{lipsum}
\usepackage{graphicx}
\usepackage{dcolumn,longtable,float,comment}
\usepackage{amsmath}

\usepackage[backend=biber, style=numeric, sorting=none, url=true, doi=true]{biblatex}
\graphicspath{ {./images/} }

    \makeatletter
\def\@fnsymbol#1{\ensuremath{\ifcase#1\or \dagger\or \dagger\dagger\or
   \mathsection\or \mathparagraph\or \|\or **\or \dagger\dagger
   \or \ddagger\ddagger \else\@ctrerr\fi}}
    \makeatother

\title{An analysis of the effects of sharing research data, code, and  preprints on citations}

\author{
 Giovanni Colavizza\thanks{giovanni.colavizza@unibo.it} \\
  University of Bologna\\
  \And
  Lauren Cadwallader \\
  PLOS \\
  \And
  Marcel LaFlamme \\
  PLOS \\
  \And
  Grégory Dozot \\
  HEIG-VD \\
  \AND
  Stéphane Lecorney \\
  HEIG-VD \\
  \And
  Daniel Rappo \\
  HEIG-VD \\
  \And
  Iain Hrynaszkiewicz\thanks{ihrynaszkiewicz@plos.org} \\
  PLOS \\
}

\begin{document}
\maketitle
\begin{abstract}

Calls to make scientific research more open have gained traction with a range of societal stakeholders. Open Science practices include but are not limited to the early sharing of results via preprints and openly sharing outputs such as data and code to make research more reproducible and extensible. Existing evidence shows that adopting Open Science practices has effects in several domains. In this study, we investigate whether adopting one or more Open Science practices leads to significantly higher citations for an associated publication, which is one form of academic impact. We use a novel dataset known as Open Science Indicators, produced by PLOS and DataSeer, which includes all PLOS publications from 2018 to 2023 as well as a comparison group sampled from the PMC Open Access Subset. In total, we analyze circa 122’000 publications. We calculate publication and author-level citation indicators and use a broad set of control variables to isolate the effect of Open Science Indicators on received citations. We show that Open Science practices are adopted to different degrees across scientific disciplines. We find that the early release of a publication as a preprint correlates with a significant positive citation advantage of about $20.2\%$ ($\pm .7$) on average. We also find that sharing data in an online repository correlates with a smaller yet still positive citation advantage of $4.3\%$ ($\pm .8$) on average. However, we do not find a significant citation advantage for sharing code. Further research is needed on additional or alternative measures of impact beyond citations. Our results are likely to be of interest to researchers, as well as publishers, research funders, and policymakers.

\end{abstract}

\section{Introduction}

Arising from a diverse set of cultural and technological projects at the turn of the twenty-first century~\cite{willinsky_unacknowledged_2005,tkacz_wikipedia_2014,moore_genealogy_2017}, contemporary calls to make scientific research more open point toward a no less diverse range of outcomes. One influential definition characterizes Open Science as ``transparent and accessible knowledge that is shared and developed through collaborative networks''~\cite{vicente-saez_open_2018}, encompassing knowledge objects or outputs as well as processes~\cite{leonelli_philosophy_2023}. Another developed by UNESCO defines Open Science as ``an inclusive construct that combines various movements and practices aiming to make multilingual scientific knowledge openly available, accessible and reusable for everyone, to increase scientific collaborations and sharing of information for the benefits of science and society, and to open the processes of scientific knowledge creation, evaluation and communication to societal actors beyond the traditional scientific community''~\cite{unesco_unesco_2021}. 

While acknowledging this diversity of ambitions, in what follows we focus on practices resulting in what UNESCO terms ``open scientific knowledge''~\cite{unesco_unesco_2021}: that is, the making of scientific publications and the materials that underpin them available to all, free of charge. These Open Science practices include but are not limited to Open Access publication; the early sharing of results, for example via the use of preprints; openly sharing outputs such as data, code, and protocols to make research more reproducible and extensible; and fostering rigor and transparency in study design, for example via study registration. While the uptake of these practices by researchers varies by field, career stage, and region, their prevalence is growing overall~\cite{serghiou_assessment_2021,menke_establishing_2022}. Drivers of this growth include publisher and funder policies, training and infrastructure support, and cultural change~\cite{robson_promoting_2021,armeni_widescale_2021}. The proliferation of policies for Open Science has led to a greater need to monitor the effects of these policies on Open Science~\cite{hrynaszkiewicz_survey_2021a}, although comprehensive solutions for measuring Open Science are lacking. Still, researchers, technology providers, research funders, institutions and publishers have begun to monitor the prevalence of Open Science practices (\url{https://open-science-monitoring.org/monitors/}). These efforts provide new evidence and data sources from which to understand if and how Open Science practices are being adopted, and to explore the extent to which these practices confer effects, impacts or benefits as a result of their adoption.

Despite their limitations~\cite{dougherty_citation_2022}, citation counts and other bibliometric indicators are frequently used as quantitative measures of research impact and quality, and play a role in research assessment activities that support researchers’ career progression, and awarding of research funding~\cite{aksnes_citations_2019}. As such, actions that can increase the potential for citations of researchers’ articles may be seen as desirable and may motivate changes in research and publishing practices. As monitoring of Open Science increases, there is also growing recognition that those who are subject to monitoring by indicators should be involved in the development of indicators~\cite{himanen_scope_2024}. Biomedical researchers have, for example, developed community consensus on the relative importance, for monitoring, of 19 different Open Science practices~\cite{cobey_community_2023}.

In this article, we contribute to an emerging strand of research assessing the impact of Open Science practices. We focus on a set of measurable Open Science practices that include data sharing, code sharing, and preprint posting. More specifically, we ask whether adopting any combination of these practices leads to a significantly higher citation impact for an associated publication when compared to similar publications for which authors have not adopted Open Science practices. We answer this question by leveraging a novel dataset known as Open Science Indicators, which is produced by the nonprofit Open Access publisher PLOS in partnership with DataSeer (\url{https://dataseer.ai})~\cite{hrynaszkiewicz_plos_2022}, and by adapting a previously released workflow to mine citation data from the PMC Open Access Subset~\cite{colavizza_citation_2020}. An important aspect of our contribution is the assessment of Open Science practices in combination, rather than individually as is usually the case in previous work.

\section{State of the Art}

There is evidence that adopting Open Science practices has effects or impacts in several domains: academic, societal, and economic~\cite{klebel_pathos_2024}. In terms of academic or research impacts, Open Science practices are associated with increased visibility and reuse, as measured for example by the diversity of citations and media attention received by Open Access articles~\cite{huang_open_2024,schultz_all_2021}. Open Science has been instrumental in accelerating progress on certain scientific problems~\cite{woelfle_open_2011}, in making results more transparent~\cite{besancon_open_2021}, and in addressing what has been termed the replication crisis in certain fields of research~\cite{opensciencecollaboration_estimating_2015}. Societal benefits identified in a systematic scoping review include enabling broader participation in research, by supporting citizen science and educational initiatives. However, evidence of societal benefits to policy, health, or trust in research is to date more limited~\cite{cole_societal_2024}. Economic benefits, identified from economic modeling studies and case studies, include cost and labor savings from Open Access and open (or FAIR) data, as well as increased innovation~\cite{fell_economic_2019,directorate-generalforresearchandinnovationeuropeancommission_costbenefit_2018}. However, there is a lack of causal evidence for and prospective studies of these benefits. Open Science practices have also been linked to negative impacts including imposing additional costs~\cite{hostler_invisible_2023}, reinforcing existing inequalities~\cite{ross-hellauer_dynamics_2022}, and homogenizing diverse research traditions~\cite{leonelli_open_2022}.

\subsection{Data and code sharing}

Researchers who adopt Open Science practices may see increased use and impact of their work, which can support career progression. Several studies examine the importance of data (and code) sharing for scientific advancement but diverge to some extent in their findings. Evidence shows that the novel combination of datasets leads to higher impact and visibility~\cite{yu_does_2024}. Several studies in specific research disciplines have found correlations between sharing research and increased citations of articles that share data~\cite{piwowar_sharing_2007,piwowar_data_2013,henneken_linking_2011}. Implementation of journal policies requiring data sharing has also been correlated with increased citations~\cite{christensen_study_2019}. Researchers can share data in several different ways but sharing data privately, upon request, and via supporting information files with publications are the most common approaches – despite being considered suboptimal~\cite{federer_longterm_2022,tedersoo_data_2021a}. Sharing research data in a public repository is considered best practice for data sharing but may require additional effort compared to other approaches~\cite{stuart_practical_2018}. However, in previous work, we found that, relative to sharing data upon request or as supporting information files, data sharing in repositories was correlated with a 25.36\% citation advantage on average~\cite{colavizza_citation_2020}.

While we can hypothesize that, similar to data sharing, code sharing would promote the reuse of published research that shares code, there is less evidence about whether code sharing is correlated with any effect on citations. Studies of a single journal~\cite{vandewalle_code_2012} or a small number of journals in a single field~\cite{kucharsky_code_2020a} have found mixed effects. A larger-scale study showed a correlation between links to methods including (but not limited to) code and increased citation, especially when links were still active~\cite{cao_rise_2023}. Another found that monthly citations of articles increased after their associated code repositories were made public~\cite{kang_papers_2023}.

\subsection{Preprints}

There is evidence for advantages in terms of visibility for peer-reviewed publications that were previously posted as preprints, as measured by increased citations and altmetrics~\cite{mckiernan_how_2016,fu_releasing_2019,fraser_relationship_2020,xie_preprint_2021}. This effect was examined in detail during the COVID-19 pandemic~\cite{fraser_evolving_2021}, when media coverage of health-related preprints also saw a significant uptick~\cite{fleerackers_unreviewed_2024}. Other forms of impact associated with preprint posting include receiving additional feedback, which research has shown to be constructive if variable in frequency~\cite{rzayeva_experiences_2023,carneiro_characterization_2023}. Studies examining the adoption of preprints by career stage have suggested that they have particular advantages for early-career researchers in terms of career development~\cite{sarabipour_value_2019,wolf_preprinting_2021}. At the same time, concerns over how preprints may introduce unvetted findings into the scientific record have pointed to the need for nuanced approaches to evidence synthesis~\cite{davidson_no_2023,zeraatkar_consistency_2022} and the communication of retractions~\cite{avissar-whiting_downstream_2022}.

\section{Methods and data}

To make this study entirely reproducible, we focus only on Open Access publications and release all of the accompanying code. We strictly follow and expand upon a published methodology~\cite{colavizza_citation_2020}. This methodology entails selecting a set of publications of interest, calculating publication and author-level citation counts using a larger Open Access collection, and modeling the effect of interest as independent variables. We use PLOS' Open Science Indicators version 5 as a starting point~\cite{hrynaszkiewicz_plos_2022,publiclibraryofscience_plos_2023}. The OSI publication count totals N = 124'274. We also use the PMC Open Access Subset, with all publications up to October 2023 included~\cite{nationallibraryofmedicine_pmc_2023}. The PMC Open Access Subset is used to calculate citation counts for publications and authors. Citation counts calculated using the PMC Open Access Subset have been shown to track global citation counts, and thus to be appropriate when the relative rather than absolute counts are of interest~\cite{colavizza_citation_2020}. Publications missing a known identifier (DOI, PubMed reference number, PMCID, or a publisher-specific ID), a publication date, and at least one reference are discarded. These often are editorials, letters, or similar article types. The final PMC Open Access Subset publication count totals M = 5'020'948. After an initial analysis, a limited amount of OSI publications are also discarded for being absent in the PMC Open Access Subset or identified as editorials or reviews (i.e., not research articles). Of the 124'376 publications in OSI, 121’999 (98.1\%) are processed, matched, and used for the modeling analysis that follows.

We use a linear model for quantifying the relative effect of Open Science Indicators on citation counts, as follows:

\paragraph{Dependent variable.} Citation counts for each publication are calculated using the full PMC Open Access Subset dataset (M publications above). Citations are based on identifiers, hence only references that include a valid ID are considered. Citations accumulated by the preprint of a publication are therefore not counted, as the preprint will have a different identifier than the published paper as included in OSI. Citations from a preprint to a published paper that is part of OSI are, however, counted. Under these constraints, we calculate total citation counts and use this as our main dependent variable. We also calculate citations given within a certain time window from publication (1, 2, and 3 years, also considering the month of publication). This is done in order to conduct a robustness check using citation counts over the same citation accrual time as the dependent variable (e.g., the three-year window for a publication published in June 2015 runs to June 2018 excluded).

\paragraph{Independent variables.} We use a set of control variables for modeling. Firstly, publication-level variables are commonly considered in similar studies~\cite{gargouri_selfselected_2010,yegros-yegros_does_2015,wang_bias_2017}. We include the year of publication, to account for citation inflation over time; the month of publication (missing values are set to a default value of 6, that is June), to account for the advantage of publications published early in the year that have more time to accrue citations; the number of authors and the total number of references (including those without a known identifier), both usually correlated with citation impact. We also use the Australian and New Zealand Standard Research Classification (ANZSRC) Fields of Research classification system at the publication level, to account for disciplinary variation in the adoption of Open Science practices. We use the broadest level provided, that of the Division, to avoid data sparsity. In the dataset, 22 divisions are found. We group the least frequent five categories into a single category, since they all belong to the Arts and Humanities. We therefore end up with 18 distinct categories that are encoded as dummy variables to account for the fact that a publication can belong to multiple categories. See Table~\ref{tab:ANZSRC} for a list of the categories used from the division-level ANZSRC Fields of Research.

\begin{table}[]
\caption{ANZSRC Fields of Research Divisions to model categories. Note that the total publication count is higher than the number of publications in OSI, since a publication can belong to more than one division.}
\label{tab:ANZSRC}
\begin{tabular}{|l|r|r|}
\hline
\textbf{ANZSRC FoR Division} & \multicolumn{1}{l|}{\textbf{Publication counts in OSI}} & \multicolumn{1}{l|}{\textbf{Category (regression model)}} \\ \hline
32 Biomedical and Clinical Sciences           & 59'377 & division\_1  \\ \hline
31 Biological Sciences                        & 35'081 & division\_2  \\ \hline
42 Health Sciences                            & 29'778 & division\_3  \\ \hline
30 Agricultural, Veterinary and Food Sciences & 8549  & division\_4  \\ \hline
46 Information and Computing Sciences         & 7704  & division\_5  \\ \hline
52 Psychology                                 & 6910  & division\_6  \\ \hline
44 Human Society                              & 5645  & division\_7  \\ \hline
40 Engineering                                & 5294  & division\_8  \\ \hline
41 Environmental Sciences                     & 5208  & division\_9  \\ \hline
34 Chemical Sciences                          & 3468  & division\_10 \\ \hline
35 Commerce, Management, Tourism and Services & 2811  & division\_11 \\ \hline
37 Earth Sciences                             & 2716  & division\_12 \\ \hline
38 Economics                                  & 1960  & division\_13 \\ \hline
51 Physical Sciences                          & 1313  & division\_14 \\ \hline
39 Education                                  & 1153  & division\_15 \\ \hline
47 Language, Communication and Culture        & 988   & division\_16 \\ \hline
49 Mathematical Sciences                      & 901   & division\_17 \\ \hline
43 History, Heritage and Archaeology          & 838   & division\_18 \\ \hline
48 Law and Legal Studies                      & 792   & division\_18 \\ \hline
33 Built Environment and Design               & 662   & division\_18 \\ \hline
36 Creative Arts and Writing                  & 459   & division\_18 \\ \hline
50 Philosophy and Religious Studies           & 399   & division\_18 \\ \hline
\end{tabular}
\end{table}

The reputation of authors before publication has also been linked to the citation success of a paper~\cite{sekara_chaperone_2018}. To control for this, we have to identify individual authors, a challenging task in itself~\cite{torvik_author_2009,lu_pubmed_2011,ferreira_brief_2012,liu_author_2014,zheng_entity_2015}. We focus on a publication-level aggregated indicator of author productivity and popularity: the mean H-index of a publication’s authors at the time of publication, calculated from the PMC Open Access Subset. While we acknowledge that the H-index has limitations and may, for instance, generate inconsistent rankings~\cite{waltman_inconsistency_2012}, in our work we do not use the H-index to this end but instead use it as a proxy for the productivity and popularity of the authors of a publication. By using the mean H-index of the authors of a publication, we further minimize the impact of errors arising from disambiguating author names~\cite{strotmann_author_2012,kim_distortive_2016}, which would have been higher if we had used measures based on individual observations such as the maximum H-index. We therefore use a simple disambiguation technique when compared to the current state of the art, and consider two author mentions to refer to the same individual if both full name and surname are found to be identical within all of the PMC Open Access Subset. We acknowledge the limitations of this method in possibly merging different authors with the same name and surname. We identify 8'481'129 seemingly distinct authors in this way.

We finally consider the following journal-level variables: if a publication is published by PLOS (any journal), and if a publication is published in PLOS ONE specifically. Given the preponderance of PLOS publications (101'366, or nearly 85\% of publications overall), and specifically PLOS ONE publications (83'843, or nearly 70\% of publications overall) in the dataset, we do not use any other journal-level variable. 

\begin{figure}[h]
\caption{Correlation plot among most variables. We see that no two variables are too highly correlated, except as expected for two alternatives for dependent variables (\texttt{n\_cit\_2} and \texttt{n\_cit\_tot)}.}\label{fig:corrplot}
\centering
\includegraphics[width=0.9\textwidth]{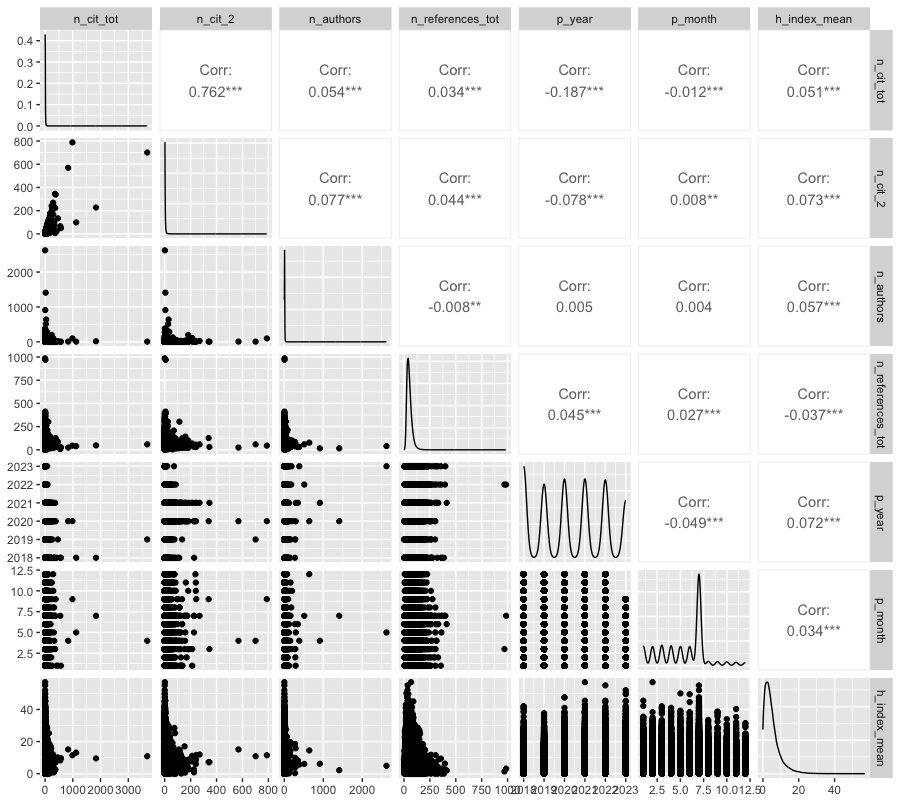}
\end{figure}

A set of descriptive statistics for the numerical variables in use is reported in Tables~\ref{tab:descr_1} and \ref{tab:descr_2}, while their correlations are illustrated in Figure~\ref{fig:corrplot}. The models we test, besides OLS linear regression and robust linear regression, include ANOVA, Tobit, and GLM with negative binomial, zero-inflated negative binomial, lognormal, and Pareto 2 family distributions. These largely support the findings using linear regression and robust linear regression, which are easier to interpret. Therefore, results from other models are omitted here and can be reproduced using the accompanying codebase. Robust linear regression results differ little from simple linear regression, as is expected given the log transformations we systematically apply on skewed numerical variables, but they are provided for comparison.

\begin{table}
\begin{center}
\caption{Descriptive statistics for the dependent variable and a set of publication and author level controls.} 
  \label{tab:descr_1}
  
\begin{tabular}{lrrrrrrr}
  
\toprule
        & n\_cit\_tot & n\_cit\_2 & n\_authors & n\_references\_tot &  p\_year & p\_month &  h\_index\_mean \\
\midrule
   Min. &       0 &     0 &       1 &              0 &  2018 &     1 &           0 \\
1st Qu. &       0 &     0 &       4 &             34 &  2019 &     3 &           2 \\
 Median &       2 &     1.0 &       6 &             46 &  2020 &     6 &           3.9 \\
   Mean &       5.1 &     2.3 &       7.1 &             51.1 &  2020 &     5.4 &           5 \\
3rd Qu. &       6 &     3 &       9 &             63 &  2022 &     7 &           6.6 \\
   Max. &    3683 &   788 &    2621 &            986 &  2023 &    12 &          57 \\
   NA's &           &         &           &                  &         &         &         425 \\
\bottomrule
\end{tabular}
\end{center}
\end{table}

\begin{table}
\begin{center}
\caption{Descriptive statistics for the OSI controls. C: Code; D: Data; Repo: Repository; P: Preprint.} 
  \label{tab:descr_2}
  
\begin{tabular}{lllllll}
\toprule
 D\_Shared &          D\_Location & Repo\_Data & C\_Generated &   C\_Shared &         C\_Location &   P\_Match \\
\midrule
N: 39'124 &         N/A: 39'657 & F: 94'715 &   N: 76'452 & N: 107'390 &       N/A: 107'443 & F: 97'235 \\
Y: 82'875 &      Online: 39'700 & T: 27'284 &   Y: 45'547 &  Y: 14'609 &     Online: 11'221 & T: 24'764 \\
          & Suppl. Info: 42'642 &           &             &            & Suppl. Info: 3'335 &           \\
\bottomrule
\end{tabular}
\end{center}
\end{table}

\section{Results}

We start by providing a brief descriptive overview of the Open Science Indicators in the target corpus and then proceed to the modeling section.

\subsection{Overview of the Open Science Indicators dataset}

As mentioned previously, the OSI dataset we use for analysis comprises 121'999 articles. The majority of the articles are published in PLOS journals, with the largest proportion originating from PLOS ONE. The remaining articles have been taken from 1232 different journals published by a range of publishers. Rates of adoption for each Open Science practice can be calculated from the dataset to give an overall impression of the degree to which Open Science is practiced. Table~\ref{tab:OS_descriptives} outlines the overall rates of adoption for the main Open Science practices in the dataset and shows that data (in a repository) and code sharing have a relatively low adoption rate across the dataset.

\begin{table}
    \caption{Descriptive statistics for the Open Science practices as measured in the OSI dataset.}
    \label{tab:OS_descriptives}
    \centering
    \begin{tabular}{lrr}
         &  Number&  \% of publications \\
         Publications&  121'999&  100\% \\
         Sharing data (anywhere)&  82'875&  68\%\\
         Sharing data (in a repository)&  27'284&  22\%\\
         Sharing data (online)&  33'786&  28\%\\
         Sharing code&  14'609&  12\%\\
         Has a preprint&  24'764&  20\%\\
    \end{tabular}
\end{table}

\begin{figure}[h]
\caption{Adoption of OSI over time. Each OSI remains adopted by a fraction of publications, but adoption grows over time.}\label{fig:OSI_time}
\centering
\includegraphics[width=0.8\textwidth]{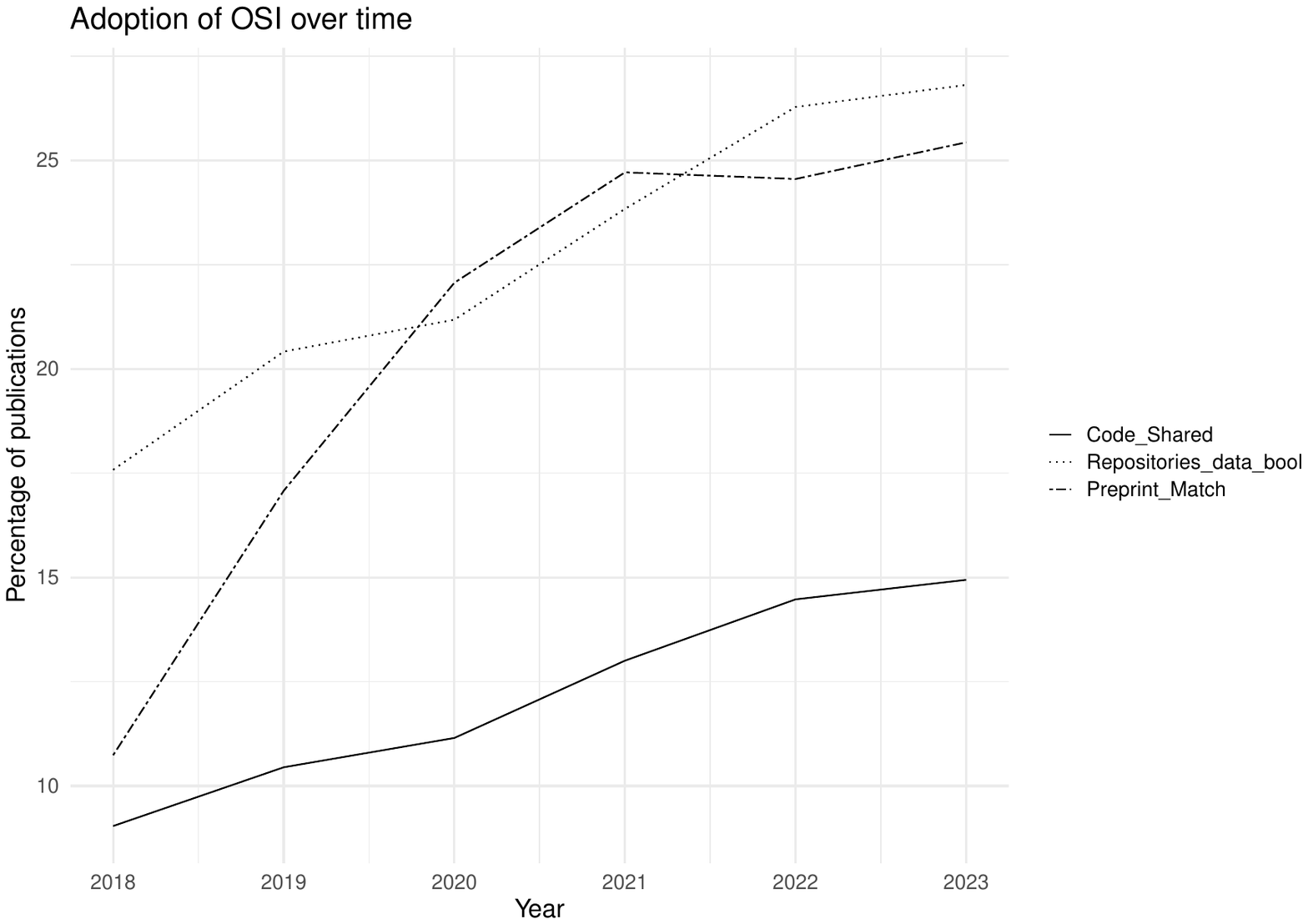}
\end{figure}

In OSI, the average rates of adoption for Open Science practices observed in the dataset have been increasing over time with changes between 5\% and 15\% from 2018 to 2023. Data sharing in any form has seen a 5\% increase from 2018 to 2023, with data sharing in repositories and online increasing by 9\% and 10\% respectively. Code sharing (out of all publications) has increased by 6\% over the same time period and preprint posting by 15\%. Whilst data and code sharing show positive trends over time, the trend for preprint posting shows a large increase between 2018 and 2019 and again from 2019 to 2020, followed by a plateauing since 2021. These trends are also seen when the PLOS cohort and the PMC Open Access Subset cohort are considered separately, although the PMC Open Access Subset cohort shows an increase in preprints in 2023 compared to 2022 which is not present in the PLOS data. We show the general adoption trends in Figure~\ref{fig:OSI_time}.

\begin{figure}[h]
\caption{Adoption of OSI by Division, as described in Table~\ref{tab:ANZSRC}. Each OSI remains adopted by a fraction of publications, but there is a wide variation across Divisions.}\label{fig:OSI_division}
\centering
\includegraphics[width=0.9\textwidth]{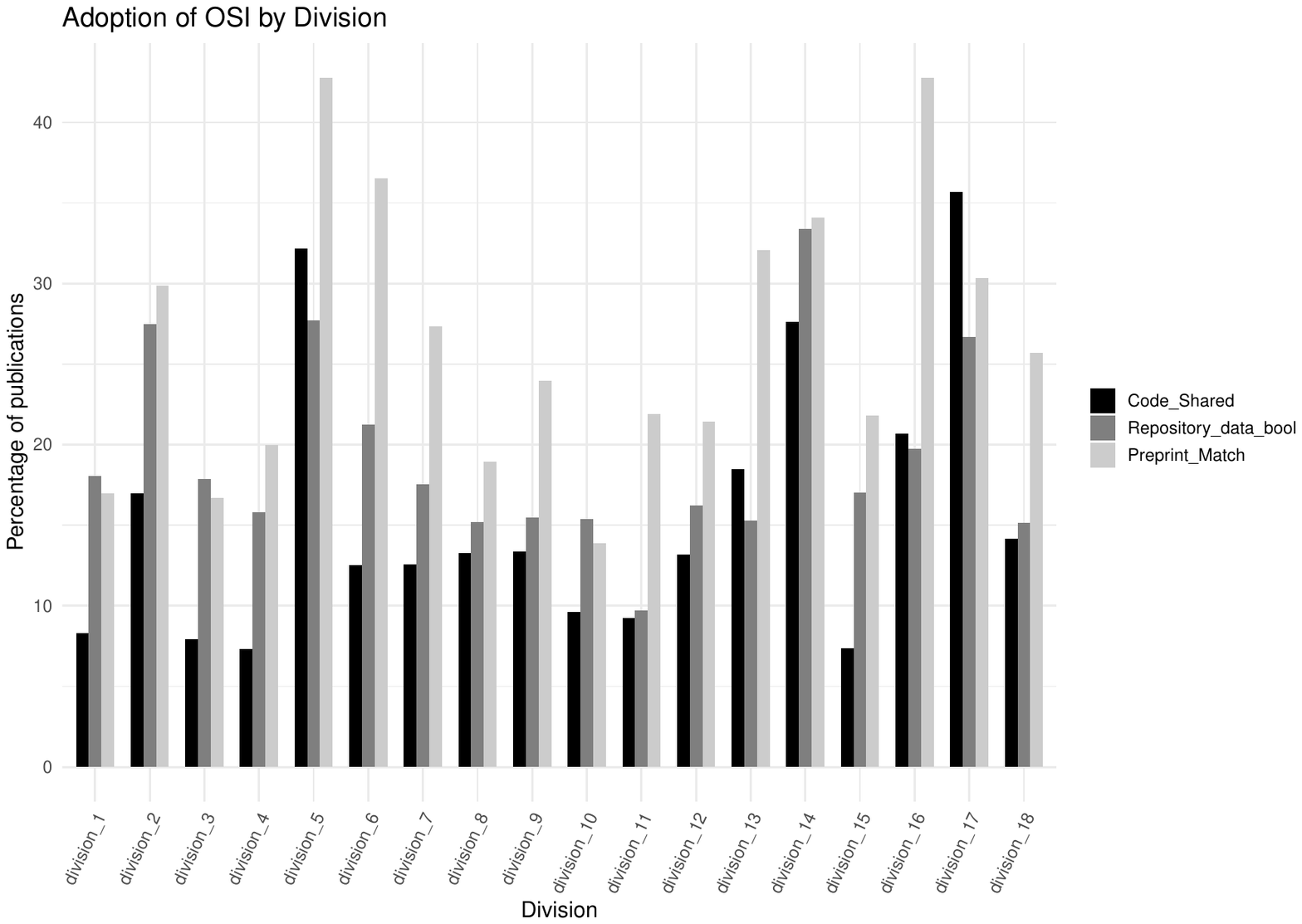}
\end{figure}

The prevalence of different Open Science practices varies by field of research (following the divisions presented in~\ref{tab:ANZSRC}). For example, both Division 3 (Health Sciences) and Division 8 (Engineering) have the lowest data sharing rate at 60\%, whilst Division 16 (Language, Communication and Culture) has the highest data sharing rate at 82\%. Similar degrees in variation are seen for the other indicators with data sharing in a repository ranging from 14\% to 43\%, code sharing from 7\% to 36\%, and preprint posting from 10\% to 33\%. Such wide variation in OSI adoption across divisions suggests that research fields face different challenges in adopting Open Science practices, and some practices may not be equally useful or relevant across fields. In Figure~\ref{fig:OSI_division}, we show trends for the main OSIs across all Divisions, as described in Table~\ref{tab:ANZSRC}.

Please refer to the PLOS' Open Science Indicators version 5 documentation for further details~\cite{hrynaszkiewicz_plos_2022,publiclibraryofscience_plos_2023}.

\subsection{Modeling}

The base model results we discuss are provided in Table~\ref{tab:model_base}. It contains the basic author, publication, and journal-level variables we discussed above. It does not contain publication-level division classification. The most complete model we discuss is instead provided in Table~\ref{tab:model_full}. Here, we use all the previous variables from the base model and add the publication-level division classification as dummy variables (division 1 to 18, see Table~\ref{tab:ANZSRC}). Several more models were tested, primarily as robustness checks, and are discussed in the Appendix. 

\begin{figure}[H]
\caption{An illustration of the assumed causal dependency graph among dependent and independent variables. We distinguish among the dependent variable and its variations (red), independent control variables (blue), and OSI control variables (green). We are interested in the total effect of OSI variables on the dependent variable (\texttt{n\_cit\_tot)}, shown by the thick black line, and in controlling for the effect of other independent variables, shown by the dotted black lines.}\label{fig:model}
\centering
\includegraphics[width=0.9\textwidth]{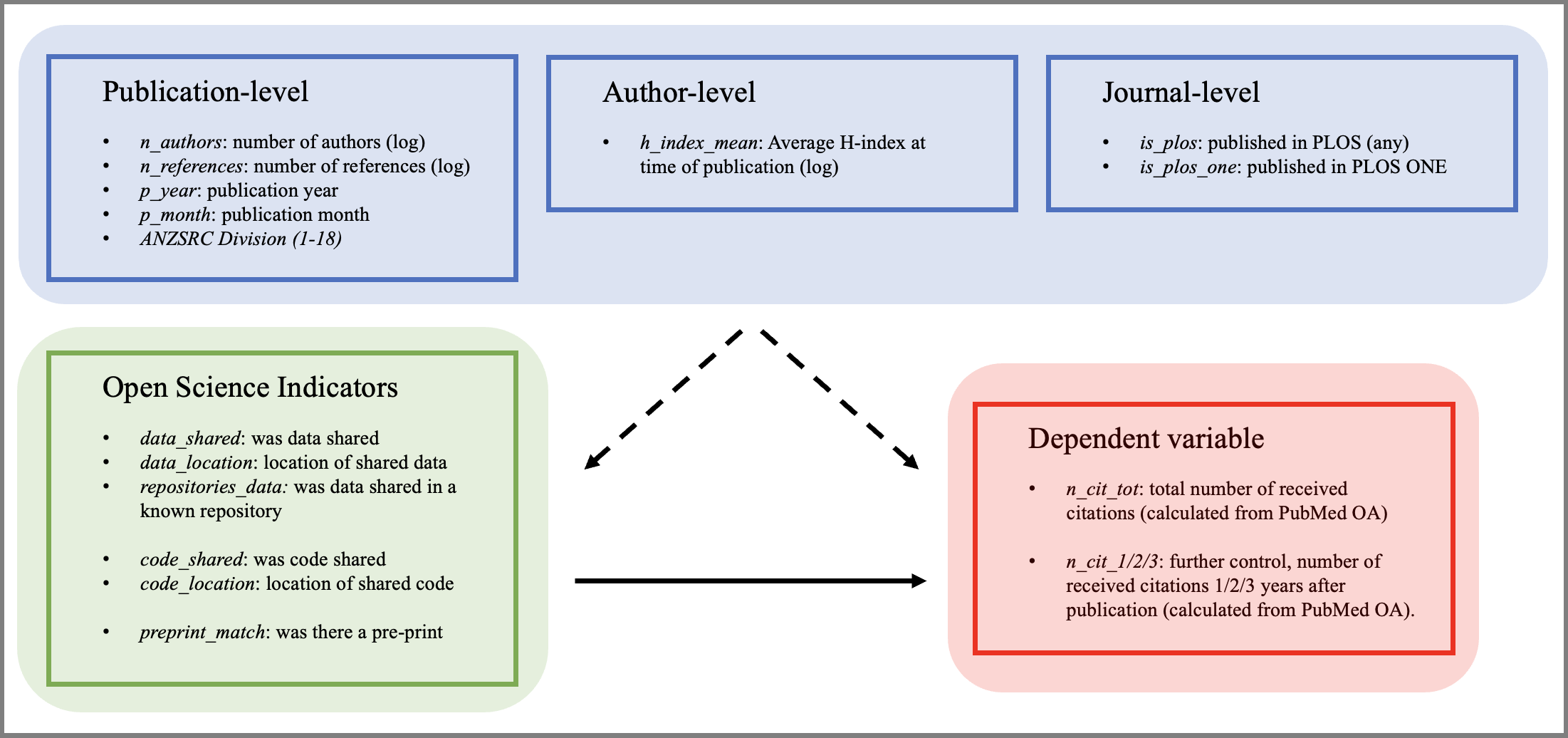}
\end{figure}

The base model is described in Equation~\ref{eq:without_division}, and the full model is described in Equation~\ref{eq:with_division}. Variable transformations are shown, numerical variables are given in Italics, and categorical variables are in regular text. Variables are grouped along lines. An illustration of the assumed causal dependency graph among variable groups is given in Figure~\ref{fig:model}. In the same figure, the variables for which we used log scaling to limit the effects of outliers are flagged as such. These include the dependent variable (\texttt{n\_cit\_tot}), which is always used on a log scale.

\begin{equation}
\begin{aligned}
\log(n\_cit\_tot + 1) = & \log(n\_authors + 1) + \log(n\_references + 1) + p\_year + p\_month + \\
             & \log(h\_index\_mean + 1) + \\
             & \text{is\_plos} + \text{is\_plos\_one} + \\
             & \text{data\_shared} + \text{data\_location} + \text{repositories\_data} + \\
             & \text{code\_shared} + \text{code\_location} + \\
             & \text{preprint\_match}
\end{aligned}
\label{eq:without_division}
\end{equation}

\begin{equation}
\begin{aligned}
\log(n\_cit\_tot + 1) = & \log(n\_authors + 1) + \log(n\_references + 1) + p\_year + p\_month + \\
             & \log(h\_index\_mean + 1) + \\
             & \text{is\_plos} + \text{is\_plos\_one} + \\
             & \text{data\_shared} + \text{data\_location} + \text{repositories\_data} + \\
             & \text{code\_shared} + \text{code\_location} + \\
             & \text{preprint\_match} + \\
             & \sum_{i=1}^{18} \text{I(division = i)}
\end{aligned}
\label{eq:with_division}
\end{equation}

Starting with the base model in Table~\ref{tab:model_base}, we provide results for an OLS model and a robust linear model as a comparison. The results are aligned and show a relatively high explained variance with the base model having $R^2 = .408$. The model shows expected trends, as previously discussed. For example, the higher the year the lower the total citation count on average (about $-30\%$ per year increase), or the higher the average H-index of the authors, the higher the citation counts of the paper (this can be interpreted as an elasticity in a log-log model, therefore a 1\% increase in the average H-index leads to a $.141\%$ increase in the number of citations, on average). More of interest to us are the OSIs. These show that there is a significant and positive effect of preprints ($20.4\%$) and of sharing data via an online repository ($3.9\%$). These percentage changes for log-linear relationships are calculated as follows: $(\exp(.186) - 1) \times 100 \approx 20.4\%$. These effects are cumulative, so a publication with both a preprint and data shared in a repository would be associated with an average citation increase of $24.3\%$. On the other hand, the OSI for code sharing did not yield a statistically significant positive citation effect. Our next question is whether these results hold when we account for the large disciplinary variations in the adoption of Open Science practices, which we assess next.

\begin{longtable}{@{\extracolsep{5pt}}lD{.}{.}{-3} D{.}{.}{-3} } 

  \caption{Results for the base model.} 
  \label{tab:model_base} 
  
\\[-1.8ex]\hline 
\hline \\[-1.8ex] 
 & \multicolumn{2}{c}{\textit{Dependent variable:}} \\ 
\cline{2-3} 
\\[-1.8ex] & \multicolumn{2}{c}{n\_cit\_tot\_log} \\ 
\\[-1.8ex] & \multicolumn{1}{c}{\textit{OLS}} & \multicolumn{1}{c}{\textit{robust}} \\ 
 & \multicolumn{1}{c}{\textit{}} & \multicolumn{1}{c}{\textit{linear}} \\ 
\\[-1.8ex] & \multicolumn{1}{c}{(1)} & \multicolumn{1}{c}{(2)}\\ 
\hline \\[-1.8ex] 
 n\_authors\_log & 0.265^{***} & 0.254^{***} \\ 
  & (0.005) & (0.004) \\ 
  & & \\ 
 n\_references\_tot\_log & 0.192^{***} & 0.198^{***} \\ 
  & (0.005) & (0.005) \\ 
  & & \\ 
 p\_year & -0.357^{***} & -0.370^{***} \\ 
  & (0.001) & (0.001) \\ 
  & & \\ 
 p\_month & -0.037^{***} & -0.039^{***} \\ 
  & (0.001) & (0.001) \\ 
  & & \\ 
 h\_index\_mean\_log & 0.141^{***} & 0.141^{***} \\ 
  & (0.003) & (0.003) \\ 
  & & \\ 
 C(is\_plos)True & 0.095^{***} & 0.107^{***} \\ 
  & (0.009) & (0.008) \\ 
  & & \\ 
 C(is\_plos\_one)True & -0.347^{***} & -0.351^{***} \\ 
  & (0.007) & (0.007) \\ 
  & & \\ 
 C(Data\_Shared)True & -0.011 & -0.012 \\ 
  & (0.034) & (0.033) \\ 
  & & \\ 
 C(Data\_Location)Online & -0.0002 & -0.001 \\ 
  & (0.034) & (0.034) \\ 
  & & \\ 
 C(Data\_Location)Supplementary Information & 0.024 & 0.027 \\ 
  & (0.034) & (0.033) \\ 
  & & \\ 
 C(Repositories\_data\_bool)True & 0.038^{***} & 0.038^{***} \\ 
  & (0.008) & (0.008) \\ 
  & & \\ 
 C(Code\_Shared)True & 0.057 & 0.064 \\ 
  & (0.107) & (0.104) \\ 
  & & \\ 
 C(Code\_Location)Online & -0.105 & -0.127 \\ 
  & (0.107) & (0.105) \\ 
  & & \\ 
 C(Code\_Location)Supplementary Information & -0.130 & -0.132 \\ 
  & (0.108) & (0.105) \\ 
  & & \\ 
 C(Preprint\_Match)True & 0.186^{***} & 0.159^{***} \\ 
  & (0.006) & (0.006) \\ 
  & & \\ 
 Constant & 720.843^{***} & 746.799^{***} \\ 
  & (2.705) & (2.638) \\ 
  & & \\ 
\hline \\[-1.8ex] 
Observations & \multicolumn{1}{c}{121,999} & \multicolumn{1}{c}{121,999} \\ 
R$^{2}$ & \multicolumn{1}{c}{0.408} &  \\ 
Adjusted R$^{2}$ & \multicolumn{1}{c}{0.408} &  \\ 
Residual Std. Error (df = 121983) & \multicolumn{1}{c}{0.775} & \multicolumn{1}{c}{0.723} \\ 
F Statistic & \multicolumn{1}{c}{5,615.550$^{***}$ (df = 15; 121983)} &  \\ 
\hline 
\hline \\[-1.8ex] 
\textit{Note:}  & \multicolumn{2}{r}{$^{*}$p$<$0.1; $^{**}$p$<$0.05; $^{***}$p$<$0.01} \\ 
\end{longtable} 

The full model in Table~\ref{tab:model_full} adds the ANZSRC divisions as 18 dummy variables. The model shows an even higher explained variance with $R^2 = 0.426$. The full model shows trends that largely confirm the results from the base model. We consolidate our estimate for the citation impact of OSI indicators as follows. We find that the early release of a publication as a preprint correlates with a significant positive citation advantage of about $20.2\%$ ($\pm .7$) on average. We also find that sharing data in an online repository is associated with a smaller yet still positive citation advantage of $4.3\%$ ($\pm .8$) on average. These effects are cumulative, so a publication with both a preprint and data shared in a repository would be associated with an average citation increase of $24.5\%$. We do not find a significant effect for sharing code, and we detect significant variations across disciplines in average citation impact. All the remaining coefficients are confirmed in sign and, with minor variation, in magnitude.

We further zoom in on individual divisions by running the base model using data points from each division only and do not find significant effects for code sharing. The effect of sharing data in an online repository varies, but is larger in two divisions (division 2, Biological Sciences, and division 8, Engineering). The effect of preprints remains consistent across divisions, with varying degrees of significance and magnitude.

\begin{longtable}{@{\extracolsep{5pt}}lD{.}{.}{-3} D{.}{.}{-3} } 

  \caption{Results for the full model.} 
  \label{tab:model_full} 
  
\\[-1.8ex]\hline 
\hline \\[-1.8ex] 
 & \multicolumn{2}{c}{\textit{Dependent variable:}} \\ 
\cline{2-3} 
\\[-1.8ex] & \multicolumn{2}{c}{n\_cit\_tot\_log} \\ 
\\[-1.8ex] & \multicolumn{1}{c}{\textit{OLS}} & \multicolumn{1}{c}{\textit{robust}} \\ 
 & \multicolumn{1}{c}{\textit{}} & \multicolumn{1}{c}{\textit{linear}} \\ 
\\[-1.8ex] & \multicolumn{1}{c}{(1)} & \multicolumn{1}{c}{(2)}\\ 
\hline \\[-1.8ex] 
 n\_authors\_log & 0.207^{***} & 0.194^{***} \\ 
  & (0.005) & (0.005) \\ 
  & & \\ 
 n\_references\_tot\_log & 0.246^{***} & 0.252^{***} \\ 
  & (0.005) & (0.005) \\ 
  & & \\ 
 p\_year & -0.357^{***} & -0.368^{***} \\ 
  & (0.001) & (0.001) \\ 
  & & \\ 
 p\_month & -0.037^{***} & -0.038^{***} \\ 
  & (0.001) & (0.001) \\ 
  & & \\ 
 h\_index\_mean\_log & 0.119^{***} & 0.120^{***} \\ 
  & (0.003) & (0.003) \\ 
  & & \\ 
 C(is\_plos)True & 0.058^{***} & 0.070^{***} \\ 
  & (0.009) & (0.008) \\ 
  & & \\ 
 C(is\_plos\_one)True & -0.304^{***} & -0.306^{***} \\ 
  & (0.007) & (0.007) \\ 
  & & \\ 
 C(Data\_Shared)True & 0.002 & -0.005 \\ 
  & (0.033) & (0.033) \\ 
  & & \\ 
 C(Data\_Location)Online & 0.010 & 0.015 \\ 
  & (0.034) & (0.033) \\ 
  & & \\ 
 C(Data\_Location)Supplementary Information & 0.020 & 0.027 \\ 
  & (0.033) & (0.033) \\ 
  & & \\ 
 C(Repositories\_data\_bool)True & 0.042^{***} & 0.041^{***} \\ 
  & (0.008) & (0.008) \\ 
  & & \\ 
 C(Code\_Shared)True & 0.099 & 0.107 \\ 
  & (0.105) & (0.103) \\ 
  & & \\ 
 C(Code\_Location)Online & -0.110 & -0.134 \\ 
  & (0.106) & (0.103) \\ 
  & & \\ 
 C(Code\_Location)Supplementary Information & -0.147 & -0.154 \\ 
  & (0.106) & (0.104) \\ 
  & & \\ 
 C(Preprint\_Match)True & 0.184^{***} & 0.155^{***} \\ 
  & (0.006) & (0.006) \\ 
  & & \\ 
 C(division\_1)True & 0.126^{***} & 0.128^{***} \\ 
  & (0.006) & (0.006) \\ 
  & & \\ 
 C(division\_2)True & 0.023^{***} & 0.031^{***} \\ 
  & (0.006) & (0.006) \\ 
  & & \\ 
 C(division\_3)True & 0.099^{***} & 0.089^{***} \\ 
  & (0.006) & (0.006) \\ 
  & & \\ 
 C(division\_4)True & 0.018^{*} & 0.030^{***} \\ 
  & (0.009) & (0.009) \\ 
  & & \\ 
 C(division\_5)True & -0.086^{***} & -0.084^{***} \\ 
  & (0.010) & (0.010) \\ 
  & & \\ 
 C(division\_6)True & -0.075^{***} & -0.087^{***} \\ 
  & (0.010) & (0.010) \\ 
  & & \\ 
 C(division\_7)True & 0.031^{***} & 0.020^{*} \\ 
  & (0.011) & (0.011) \\ 
  & & \\ 
 C(division\_8)True & -0.186^{***} & -0.172^{***} \\ 
  & (0.011) & (0.011) \\ 
  & & \\ 
 C(division\_9)True & -0.301^{***} & -0.294^{***} \\ 
  & (0.012) & (0.011) \\ 
  & & \\ 
 C(division\_10)True & 0.035^{**} & 0.045^{***} \\ 
  & (0.014) & (0.013) \\ 
  & & \\ 
 C(division\_11)True & -0.059^{***} & -0.046^{***} \\ 
  & (0.015) & (0.015) \\ 
  & & \\ 
 C(division\_12)True & -0.235^{***} & -0.238^{***} \\ 
  & (0.015) & (0.015) \\ 
  & & \\ 
 C(division\_13)True & -0.137^{***} & -0.133^{***} \\ 
  & (0.018) & (0.018) \\ 
  & & \\ 
 C(division\_14)True & -0.147^{***} & -0.138^{***} \\ 
  & (0.021) & (0.021) \\ 
  & & \\ 
 C(division\_15)True & -0.037 & -0.017 \\ 
  & (0.023) & (0.023) \\ 
  & & \\ 
 C(division\_16)True & -0.057^{**} & -0.064^{***} \\ 
  & (0.025) & (0.024) \\ 
  & & \\ 
 C(division\_17)True & -0.173^{***} & -0.164^{***} \\ 
  & (0.026) & (0.025) \\ 
  & & \\ 
 C(division\_18)True & -0.111^{***} & -0.112^{***} \\ 
  & (0.015) & (0.014) \\ 
  & & \\ 
 Constant & 721.097^{***} & 744.009^{***} \\ 
  & (2.698) & (2.635) \\ 
  & & \\ 
\hline \\[-1.8ex] 
Observations & \multicolumn{1}{c}{121,999} & \multicolumn{1}{c}{121,999} \\ 
R$^{2}$ & \multicolumn{1}{c}{0.426} &  \\ 
Adjusted R$^{2}$ & \multicolumn{1}{c}{0.426} &  \\ 
Residual Std. Error (df = 121965) & \multicolumn{1}{c}{0.764} & \multicolumn{1}{c}{0.714} \\ 
F Statistic & \multicolumn{1}{c}{2,739.270$^{***}$ (df = 33; 121965)} &  \\ 
\hline 
\hline \\[-1.8ex] 
\textit{Note:}  & \multicolumn{2}{r}{$^{*}$p$<$0.1; $^{**}$p$<$0.05; $^{***}$p$<$0.01} \\ 
\end{longtable}

\section{Discussion}

This study offers a comprehensive analysis of the citation impact of Open Science practices, drawing on a dataset of about 122'000 research articles and using both descriptive and regression analysis. Our findings reveal a consistent citation advantage for articles whose authors adopted Open Science practices, including data sharing in online repositories and preprint posting. This correlation suggests that Open Science practices may significantly enhance the visibility and academic impact of research findings. However, the Open Science practice of sharing code does not seem to lead to a citation advantage in our sample.

\subsection{Limitations}
Several limitations of our study should be acknowledged. First, while our dataset is extensive, it is heavily weighted toward publications by the Open Access publisher PLOS, and as such it may not fully capture the diversity of research across all fields, potentially limiting the generalizability of our findings. Furthermore, PLOS champions Open Science practices, and the stance that a publisher takes in this regard may have an influence on the observed effects. In particular, PLOS requires all authors, with limited exceptions, to share the research data supporting their articles as a condition of publication, with the use of data repositories as the preferred approach. This is reflected in the higher overall rates of data sharing, and higher rates of data repository use in PLOS articles compared to comparators in the OSI dataset. As data sharing is the norm in PLOS articles and as the use of repositories is not uncommon, a citation advantage for the use of data repositories may be smaller in PLOS articles compared to non-PLOS articles. Posting preprints, however, is an optional practice for researchers publishing with PLOS and most other journals. Code sharing, similarly, is optional in most of the journals in our sample, with rare exceptions such as PLOS Computational Biology, where this practice is mandatory~\cite{cadwallader_advancing_2022}.

Additionally, the observational nature of our study precludes definitive conclusions about causality. The observed citation advantage might be influenced by other factors not accounted for in our analysis, such as the intrinsic quality of the research or access to research funding. 

\subsection{Extension of previous research}
The model-explained variance in our results is globally high with respect to similar studies. For instance, there is previous work showing a positive correlation between citation and altmetric impact of publications, and the posting of preprints~\cite{mckiernan_how_2016,fu_releasing_2019,fraser_relationship_2020,xie_preprint_2021}. The extent of the citation advantage, previously found to be as much as fivefold, is known to vary according to the timing of preprint posting, the discipline, and the preprint server used, among other factors. The smaller magnitude of the effect we find relative to previous studies may relate to the broader range of preprint servers that our sample considers.

Using similar methods to ours, previous work also found a correlation between articles that include statements linking to data in a repository and a citation advantage of up to 25\%~\cite{colavizza_citation_2020}. We confirm this finding in our study, finding a positive correlation between sharing data in a repository and citation impact. Yet the effect we find is considerably smaller in magnitude. This might be caused by the smaller and more uniform dataset that we use here, which includes all PLOS publications and a smaller comparator set, while this previous work used all PLOS and BMC articles and a dataset of over half a million publications. Other studies have also found a positive citation impact of the use of discipline-specific repositories~\cite{piwowar_sharing_2007,piwowar_data_2013,henneken_linking_2011}.

While previous work~\cite{vandewalle_code_2012,cao_rise_2023,kang_papers_2023} has found as much as a threefold citation advantage for code sharing, we did not confirm this finding in our sample. Following~\cite{escamilla_rise_2022}, it is possible that outside of fields like computer science authors are more likely to cite or link to shared code directly rather than citing the research paper with which it was associated. Another possible reason could be that the quality of the description of the code, rather than its availability, impacts whether or not a research paper is cited, as was suggested in previous work on model papers~\cite{janssen_code_2020}. Without further research into code sharing and citation practices it is difficult to explain why there is a lack of significant findings for code.

\setcounter{footnote}{0} 

\subsection{Implications for future research}
Our data and code are shared openly to enable independent replication of our results and extension of our findings as larger or different, comparable sources of data on the adoption of Open Science practices become available. This includes future versions of the PLOS OSI dataset, as well as outputs from other Open Science monitoring initiatives, such as the French Open Science Monitor~\footnote{\url{https://data.enseignementsup-recherche.gouv.fr/explore/dataset/open-access-monitor-france}.} or OpenAIRE~\footnote{\url{https://monitor.openaire.eu}.}.

As Open Science practices and policies continue to develop, future research could explore longitudinal changes in citation patterns. Further studies could also investigate the relationship between additional Open Science practices and citation impact, extending our understanding of how different aspects of openness contribute to research visibility. Moreover, it would be valuable to examine the impact of Open Science practices on other domains of research dissemination and engagement, such as open commons (e.g., Wikipedia), public policy influence, collaboration networks, and public engagement. We might hypothesize, for example, that non-citation measures of impact -- such as forks and downloads -- may be more relevant for the sharing of code and software. Contemporary calls for the reform of research assessment (such as \url{https://coara.eu}) emphasize valuing more diverse research outputs and contributions, as well as more diverse measures of impact. These developments underscore the importance of future research exploring the association of Open Science practices with effects other than citations.

\section{Conclusion}

In summary, our study contributes to the growing body of literature on the effects or impacts of Open Science by quantifying the citation impact of data sharing, code sharing, and preprint posting. Our results could be readily extended with additional data on Open Science practices detected in a larger sample of non-PLOS Open Access publications. We advocate for further empirical research to build on these findings, particularly work that focuses on causal mechanisms, discipline-specific effects, and broader impacts beyond citation metrics.

\section*{Data and Code Availability}

OSI dataset: \url{https://doi.org/10.6084/m9.figshare.21687686.v5}

Code (GitHub): \url{https://github.com/MediaComem/das-public}

Data (Zenodo): \url{https://zenodo.org/doi/10.5281/zenodo.10134811}

\section*{Funding}

PLOS provided funding for data acquisition, modeling, and analysis, and had a role in the study design, analysis, and preparation of the manuscript. PLOS also provided support in the form of salaries for authors LC, ML, and IH.

\section*{Competing interests}

Three of the authors (LC, ML, and IH) were at the time of publication employed by PLOS, the publisher of PLOS ONE. This does not alter our adherence to PLOS ONE policies on sharing data and materials.

\section*{Authors’ contributions}

\begin{itemize}
    \item GC: Conceptualization, Data curation, Formal analysis, Investigation, Methodology, Software, Supervision, Writing – original draft, Writing – review \& editing.
    \item LC: Formal Analysis, Visualization, Writing – original draft.
    \item ML: Writing – original draft, Writing – review \& editing.
    \item GD: Data curation, Software, Writing – review \& editing.
    \item SL: Supervision, Writing – review \& editing.
    \item DR: Supervision, Writing – review \& editing.
    \item IH: Conceptualization, Funding acquisition, Methodology, Project administration, Resources, Supervision, Writing – original draft, Writing – review \& editing.
\end{itemize}

\section*{Acknowledgements}

We thank Tim Vines, Scott Kerr, Souad McIntosh, and the team at DataSeer for their collaboration in enabling the OSI dataset to be used for this analysis. We also thank Ross Gray at PLOS for reviewing the data and code from our study.

\section*{Appendix}

We show in this Appendix results for a few more models in order to further confirm our results. Firstly, a base model adding code generated as a variable shows a small yet significant negative effect related to it (Table \ref{tab:model_code_gen}). This effect goes away when controlling for disciplines, therefore we consider it spurious. When considering OSI interactions (Table \ref{tab:model_interactions}), we find a further negative effect provided by code generated and code shared. This surprising result may be an artifact of the dataset, that we are unsure how to explain. Next, we show how different preprint servers are associated with varying degrees of citation impact (Table \ref{tab:model_preprint_servers}). Lastly, we check a full model using as dependent variables the citation counts up to 1 year after publication (Table \ref{tab:model_1y}). We still find the same results as using the full citation counts, albeit with a smaller magnitude as expected.

\begin{longtable}{@{\extracolsep{5pt}}lD{.}{.}{-3} D{.}{.}{-3} } 

  \caption{Results for the base model with code generated OSI.} 
  \label{tab:model_code_gen}
  
\\[-1.8ex]\hline 
\hline \\[-1.8ex] 
 & \multicolumn{2}{c}{\textit{Dependent variable:}} \\ 
\cline{2-3} 
\\[-1.8ex] & \multicolumn{2}{c}{n\_cit\_tot\_log} \\ 
\\[-1.8ex] & \multicolumn{1}{c}{\textit{OLS}} & \multicolumn{1}{c}{\textit{robust}} \\ 
 & \multicolumn{1}{c}{\textit{}} & \multicolumn{1}{c}{\textit{linear}} \\ 
\\[-1.8ex] & \multicolumn{1}{c}{(1)} & \multicolumn{1}{c}{(2)}\\ 
\hline \\[-1.8ex] 
 n\_authors\_log & 0.266^{***} & 0.255^{***} \\ 
  & (0.005) & (0.004) \\ 
  & & \\ 
 n\_references\_tot\_log & 0.195^{***} & 0.200^{***} \\ 
  & (0.005) & (0.005) \\ 
  & & \\ 
 p\_year & -0.357^{***} & -0.370^{***} \\ 
  & (0.001) & (0.001) \\ 
  & & \\ 
 p\_month & -0.037^{***} & -0.039^{***} \\ 
  & (0.001) & (0.001) \\ 
  & & \\ 
 h\_index\_mean\_log & 0.141^{***} & 0.142^{***} \\ 
  & (0.003) & (0.003) \\ 
  & & \\ 
 C(is\_plos)True & 0.095^{***} & 0.106^{***} \\ 
  & (0.009) & (0.008) \\ 
  & & \\ 
 C(is\_plos\_one)True & -0.348^{***} & -0.352^{***} \\ 
  & (0.007) & (0.007) \\ 
  & & \\ 
 C(Data\_Shared)True & -0.005 & -0.007 \\ 
  & (0.034) & (0.033) \\ 
  & & \\ 
 C(Data\_Location)Online & -0.003 & -0.003 \\ 
  & (0.034) & (0.034) \\ 
  & & \\ 
 C(Data\_Location)Supplementary Information & 0.020 & 0.024 \\ 
  & (0.034) & (0.033) \\ 
  & & \\ 
 C(Repositories\_data\_bool)True & 0.041^{***} & 0.040^{***} \\ 
  & (0.008) & (0.008) \\ 
  & & \\ 
 C(Code\_Generated)True & -0.022^{***} & -0.017^{***} \\ 
  & (0.005) & (0.005) \\ 
  & & \\ 
 C(Code\_Shared)True & 0.070 & 0.075 \\ 
  & (0.107) & (0.104) \\ 
  & & \\ 
 C(Code\_Location)Online & -0.111 & -0.132 \\ 
  & (0.107) & (0.105) \\ 
  & & \\ 
 C(Code\_Location)Supplementary Information & -0.137 & -0.138 \\ 
  & (0.108) & (0.105) \\ 
  & & \\ 
 C(Preprint\_Match)True & 0.188^{***} & 0.160^{***} \\ 
  & (0.006) & (0.006) \\ 
  & & \\ 
 Constant & 720.947^{***} & 746.881^{***} \\ 
  & (2.705) & (2.638) \\ 
  & & \\ 
\hline \\[-1.8ex] 
Observations & \multicolumn{1}{c}{121,999} & \multicolumn{1}{c}{121,999} \\ 
R$^{2}$ & \multicolumn{1}{c}{0.409} &  \\ 
Adjusted R$^{2}$ & \multicolumn{1}{c}{0.408} &  \\ 
Residual Std. Error (df = 121982) & \multicolumn{1}{c}{0.775} & \multicolumn{1}{c}{0.723} \\ 
F Statistic & \multicolumn{1}{c}{5,266.481$^{***}$ (df = 16; 121982)} &  \\ 
\hline 
\hline \\[-1.8ex] 
\textit{Note:}  & \multicolumn{2}{r}{$^{*}$p$<$0.1; $^{**}$p$<$0.05; $^{***}$p$<$0.01} \\ 
\end{longtable}

\begin{longtable}{@{\extracolsep{5pt}}lD{.}{.}{-3} D{.}{.}{-3} } 

  \caption{Results for the base model with interactions among OSI.} 
  \label{tab:model_interactions}  
\\[-1.8ex]\hline 
\hline \\[-1.8ex] 
 & \multicolumn{2}{c}{\textit{Dependent variable:}} \\ 
\cline{2-3} 
\\[-1.8ex] & \multicolumn{2}{c}{n\_cit\_tot\_log} \\ 
\\[-1.8ex] & \multicolumn{1}{c}{\textit{OLS}} & \multicolumn{1}{c}{\textit{robust}} \\ 
 & \multicolumn{1}{c}{\textit{}} & \multicolumn{1}{c}{\textit{linear}} \\ 
\\[-1.8ex] & \multicolumn{1}{c}{(1)} & \multicolumn{1}{c}{(2)}\\ 
\hline \\[-1.8ex] 
 n\_authors\_log & 0.266^{***} & 0.254^{***} \\ 
  & (0.005) & (0.004) \\ 
  & & \\ 
 n\_references\_tot\_log & 0.195^{***} & 0.200^{***} \\ 
  & (0.005) & (0.005) \\ 
  & & \\ 
 p\_year & -0.357^{***} & -0.370^{***} \\ 
  & (0.001) & (0.001) \\ 
  & & \\ 
 p\_month & -0.037^{***} & -0.039^{***} \\ 
  & (0.001) & (0.001) \\ 
  & & \\ 
 h\_index\_mean\_log & 0.141^{***} & 0.142^{***} \\ 
  & (0.003) & (0.003) \\ 
  & & \\ 
 C(is\_plos)True & 0.096^{***} & 0.107^{***} \\ 
  & (0.009) & (0.008) \\ 
  & & \\ 
 C(is\_plos\_one)True & -0.350^{***} & -0.352^{***} \\ 
  & (0.007) & (0.007) \\ 
  & & \\ 
 C(Data\_Shared)True & -0.007 & -0.009 \\ 
  & (0.034) & (0.033) \\ 
  & & \\ 
 C(Data\_Location)Online & -0.003 & -0.003 \\ 
  & (0.034) & (0.034) \\ 
  & & \\ 
 C(Data\_Location)Supplementary Information & 0.022 & 0.025 \\ 
  & (0.034) & (0.033) \\ 
  & & \\ 
 C(Repositories\_data\_bool)True & 0.044^{***} & 0.041^{***} \\ 
  & (0.009) & (0.009) \\ 
  & & \\ 
 C(Preprint\_Match)True & 0.190^{***} & 0.160^{***} \\ 
  & (0.007) & (0.007) \\ 
  & & \\ 
 C(Code\_Generated)True & -0.015^{***} & -0.010^{*} \\ 
  & (0.005) & (0.005) \\ 
  & & \\ 
 C(Code\_Shared)True & 0.144 & 0.147 \\ 
  & (0.108) & (0.105) \\ 
  & & \\ 
 C(Code\_Location)Online & -0.120 & -0.142 \\ 
  & (0.107) & (0.105) \\ 
  & & \\ 
 C(Code\_Location)Supplementary Information & -0.166 & -0.166 \\ 
  & (0.108) & (0.105) \\ 
  & & \\ 
 C(Repositories\_data\_bool)True:C(Preprint\_Match)True & -0.007 & 0.002 \\ 
  & (0.012) & (0.012) \\ 
  & & \\ 
 C(Code\_Generated)True:C(Code\_Shared)True & -0.079^{***} & -0.077^{***} \\ 
  & (0.017) & (0.017) \\ 
  & & \\ 
 Constant & 720.952^{***} & 746.876^{***} \\ 
  & (2.705) & (2.638) \\ 
  & & \\ 
\hline \\[-1.8ex] 
Observations & \multicolumn{1}{c}{121,999} & \multicolumn{1}{c}{121,999} \\ 
R$^{2}$ & \multicolumn{1}{c}{0.409} &  \\ 
Adjusted R$^{2}$ & \multicolumn{1}{c}{0.409} &  \\ 
Residual Std. Error (df = 121980) & \multicolumn{1}{c}{0.775} & \multicolumn{1}{c}{0.723} \\ 
F Statistic & \multicolumn{1}{c}{4,683.261$^{***}$ (df = 18; 121980)} &  \\ 
\hline 
\hline \\[-1.8ex] 
\textit{Note:}  & \multicolumn{2}{r}{$^{*}$p$<$0.1; $^{**}$p$<$0.05; $^{***}$p$<$0.01} \\ 
\end{longtable}

\begin{longtable}{@{\extracolsep{5pt}}lD{.}{.}{-3} D{.}{.}{-3} } 

  \caption{Results for the base model with preprint servers (considering only those mentioned in 500 or more publications part of the dataset).} 
  \label{tab:model_preprint_servers} 
\\[-1.8ex]\hline 
\hline \\[-1.8ex] 
 & \multicolumn{2}{c}{\textit{Dependent variable:}} \\ 
\cline{2-3} 
\\[-1.8ex] & \multicolumn{2}{c}{n\_cit\_tot\_log} \\ 
\\[-1.8ex] & \multicolumn{1}{c}{\textit{OLS}} & \multicolumn{1}{c}{\textit{robust}} \\ 
 & \multicolumn{1}{c}{\textit{}} & \multicolumn{1}{c}{\textit{linear}} \\ 
\\[-1.8ex] & \multicolumn{1}{c}{(1)} & \multicolumn{1}{c}{(2)}\\ 
\hline \\[-1.8ex] 
 n\_authors\_log & 0.259^{***} & 0.250^{***} \\ 
  & (0.005) & (0.004) \\ 
  & & \\ 
 n\_references\_tot\_log & 0.201^{***} & 0.204^{***} \\ 
  & (0.005) & (0.005) \\ 
  & & \\ 
 p\_year & -0.360^{***} & -0.372^{***} \\ 
  & (0.001) & (0.001) \\ 
  & & \\ 
 p\_month & -0.037^{***} & -0.039^{***} \\ 
  & (0.001) & (0.001) \\ 
  & & \\ 
 h\_index\_mean\_log & 0.143^{***} & 0.143^{***} \\ 
  & (0.003) & (0.003) \\ 
  & & \\ 
 C(is\_plos)True & 0.094^{***} & 0.106^{***} \\ 
  & (0.009) & (0.009) \\ 
  & & \\ 
 C(is\_plos\_one)True & -0.344^{***} & -0.346^{***} \\ 
  & (0.007) & (0.007) \\ 
  & & \\ 
 C(Data\_Shared)True & -0.010 & -0.013 \\ 
  & (0.034) & (0.033) \\ 
  & & \\ 
 C(Data\_Location)Online & -0.001 & 0.002 \\ 
  & (0.034) & (0.034) \\ 
  & & \\ 
 C(Data\_Location)Supplementary Information & 0.027 & 0.030 \\ 
  & (0.034) & (0.033) \\ 
  & & \\ 
 C(Repositories\_data\_bool)True & 0.043^{***} & 0.041^{***} \\ 
  & (0.009) & (0.008) \\ 
  & & \\ 
 C(Code\_Shared)True & 0.060 & 0.059 \\ 
  & (0.107) & (0.105) \\ 
  & & \\ 
 C(Code\_Location)Online & -0.107 & -0.119 \\ 
  & (0.108) & (0.105) \\ 
  & & \\ 
 C(Code\_Location)Supplementary Information & -0.131 & -0.125 \\ 
  & (0.108) & (0.106) \\ 
  & & \\ 
 C(Preprint\_Match)True & 0.689^{**} & 0.264 \\ 
  & (0.345) & (0.338) \\ 
  & & \\ 
 C(Preprint\_Server)bioRxiv & 0.189^{***} & 0.191^{***} \\ 
  & (0.027) & (0.027) \\ 
  & & \\ 
 C(Preprint\_Server)Journal of Medical Internet Research & 0.517^{***} & 0.480^{***} \\ 
  & (0.038) & (0.037) \\ 
  & & \\ 
 C(Preprint\_Server)medRxiv & 0.470^{***} & 0.381^{***} \\ 
  & (0.030) & (0.029) \\ 
  & & \\ 
 C(Preprint\_Server)N/A & 0.721^{**} & 0.309 \\ 
  & (0.346) & (0.339) \\ 
  & & \\ 
 C(Preprint\_Server)Protocols.io & -0.047 & -0.040 \\ 
  & (0.043) & (0.042) \\ 
  & & \\ 
 C(Preprint\_Server)PsyArXiv & 0.181^{***} & 0.139^{***} \\ 
  & (0.039) & (0.038) \\ 
  & & \\ 
 C(Preprint\_Server)Research Square & 0.191^{***} & 0.194^{***} \\ 
  & (0.029) & (0.028) \\ 
  & & \\ 
 Constant & 726.760^{***} & 751.681^{***} \\ 
  & (2.754) & (2.695) \\ 
  & & \\ 
\hline \\[-1.8ex] 
Observations & \multicolumn{1}{c}{120,195} & \multicolumn{1}{c}{120,195} \\ 
R$^{2}$ & \multicolumn{1}{c}{0.413} &  \\ 
Adjusted R$^{2}$ & \multicolumn{1}{c}{0.413} &  \\ 
Residual Std. Error (df = 120172) & \multicolumn{1}{c}{0.772} & \multicolumn{1}{c}{0.721} \\ 
F Statistic & \multicolumn{1}{c}{3,838.518$^{***}$ (df = 22; 120172)} &  \\ 
\hline 
\hline \\[-1.8ex] 
\textit{Note:}  & \multicolumn{2}{r}{$^{*}$p$<$0.1; $^{**}$p$<$0.05; $^{***}$p$<$0.01} \\ 
\end{longtable}

\begin{longtable}{@{\extracolsep{5pt}}lD{.}{.}{-3} D{.}{.}{-3} } 

  \caption{Results for the full model with dependent variable as citation data for 1 year from publication.} 
  \label{tab:model_1y} 
\\[-1.8ex]\hline 
\hline \\[-1.8ex] 
 & \multicolumn{2}{c}{\textit{Dependent variable:}} \\ 
\cline{2-3} 
\\[-1.8ex] & \multicolumn{2}{c}{n\_cit\_1\_log} \\ 
\\[-1.8ex] & \multicolumn{1}{c}{\textit{OLS}} & \multicolumn{1}{c}{\textit{robust}} \\ 
 & \multicolumn{1}{c}{\textit{}} & \multicolumn{1}{c}{\textit{linear}} \\ 
\\[-1.8ex] & \multicolumn{1}{c}{(1)} & \multicolumn{1}{c}{(2)}\\ 
\hline \\[-1.8ex] 
 n\_authors\_log & 0.131^{***} & 0.109^{***} \\ 
  & (0.004) & (0.003) \\ 
  & & \\ 
 n\_references\_tot\_log & 0.113^{***} & 0.117^{***} \\ 
  & (0.004) & (0.004) \\ 
  & & \\ 
 p\_year & 0.020^{***} & 0.015^{***} \\ 
  & (0.001) & (0.001) \\ 
  & & \\ 
 p\_month & -0.013^{***} & -0.013^{***} \\ 
  & (0.001) & (0.001) \\ 
  & & \\ 
 h\_index\_mean\_log & 0.052^{***} & 0.050^{***} \\ 
  & (0.003) & (0.002) \\ 
  & & \\ 
 C(is\_plos)True & 0.129^{***} & 0.149^{***} \\ 
  & (0.007) & (0.006) \\ 
  & & \\ 
 C(is\_plos\_one)True & -0.303^{***} & -0.302^{***} \\ 
  & (0.006) & (0.005) \\ 
  & & \\ 
 C(Data\_Shared)True & 0.015 & 0.015 \\ 
  & (0.025) & (0.023) \\ 
  & & \\ 
 C(Data\_Location)Online & 0.007 & 0.008 \\ 
  & (0.025) & (0.023) \\ 
  & & \\ 
 C(Data\_Location)Supplementary Information & 0.003 & 0.005 \\ 
  & (0.025) & (0.023) \\ 
  & & \\ 
 C(Repositories\_data\_bool)True & 0.012^{*} & 0.011^{*} \\ 
  & (0.006) & (0.006) \\ 
  & & \\ 
 C(Code\_Shared)True & 0.084 & 0.075 \\ 
  & (0.081) & (0.075) \\ 
  & & \\ 
 C(Code\_Location)Online & -0.091 & -0.092 \\ 
  & (0.081) & (0.075) \\ 
  & & \\ 
 C(Code\_Location)Supplementary Information & -0.101 & -0.087 \\ 
  & (0.082) & (0.075) \\ 
  & & \\ 
 C(Preprint\_Match)True & 0.133^{***} & 0.099^{***} \\ 
  & (0.004) & (0.004) \\ 
  & & \\ 
 C(division\_1)True & 0.047^{***} & 0.049^{***} \\ 
  & (0.004) & (0.004) \\ 
  & & \\ 
 C(division\_2)True & 0.019^{***} & 0.029^{***} \\ 
  & (0.005) & (0.004) \\ 
  & & \\ 
 C(division\_3)True & 0.003 & -0.001 \\ 
  & (0.005) & (0.004) \\ 
  & & \\ 
 C(division\_4)True & -0.015^{**} & -0.007 \\ 
  & (0.007) & (0.006) \\ 
  & & \\ 
 C(division\_5)True & -0.037^{***} & -0.030^{***} \\ 
  & (0.008) & (0.007) \\ 
  & & \\ 
 C(division\_6)True & -0.041^{***} & -0.044^{***} \\ 
  & (0.008) & (0.007) \\ 
  & & \\ 
 C(division\_7)True & 0.006 & 0.0003 \\ 
  & (0.009) & (0.008) \\ 
  & & \\ 
 C(division\_8)True & -0.073^{***} & -0.054^{***} \\ 
  & (0.009) & (0.008) \\ 
  & & \\ 
 C(division\_9)True & -0.119^{***} & -0.104^{***} \\ 
  & (0.009) & (0.008) \\ 
  & & \\ 
 C(division\_10)True & 0.045^{***} & 0.056^{***} \\ 
  & (0.010) & (0.009) \\ 
  & & \\ 
 C(division\_11)True & -0.026^{**} & -0.019^{*} \\ 
  & (0.012) & (0.011) \\ 
  & & \\ 
 C(division\_12)True & -0.056^{***} & -0.057^{***} \\ 
  & (0.012) & (0.011) \\ 
  & & \\ 
 C(division\_13)True & -0.037^{**} & -0.030^{**} \\ 
  & (0.014) & (0.013) \\ 
  & & \\ 
 C(division\_14)True & -0.072^{***} & -0.062^{***} \\ 
  & (0.016) & (0.015) \\ 
  & & \\ 
 C(division\_15)True & -0.024 & -0.012 \\ 
  & (0.018) & (0.017) \\ 
  & & \\ 
 C(division\_16)True & 0.010 & 0.0002 \\ 
  & (0.019) & (0.018) \\ 
  & & \\ 
 C(division\_17)True & -0.056^{***} & -0.045^{**} \\ 
  & (0.020) & (0.018) \\ 
  & & \\ 
 C(division\_18)True & -0.032^{***} & -0.033^{***} \\ 
  & (0.011) & (0.010) \\ 
  & & \\ 
 Constant & -40.101^{***} & -29.688^{***} \\ 
  & (2.385) & (2.198) \\ 
  & & \\ 
\hline \\[-1.8ex] 
Observations & \multicolumn{1}{c}{106,733} & \multicolumn{1}{c}{106,733} \\ 
R$^{2}$ & \multicolumn{1}{c}{0.120} &  \\ 
Adjusted R$^{2}$ & \multicolumn{1}{c}{0.120} &  \\ 
Residual Std. Error (df = 106699) & \multicolumn{1}{c}{0.542} & \multicolumn{1}{c}{0.522} \\ 
F Statistic & \multicolumn{1}{c}{442.350$^{***}$ (df = 33; 106699)} &  \\ 
\hline 
\hline \\[-1.8ex] 
\textit{Note:}  & \multicolumn{2}{r}{$^{*}$p$<$0.1; $^{**}$p$<$0.05; $^{***}$p$<$0.01} \\ 
\end{longtable}


\printbibliography

\end{document}